\documentclass[reqno]{amsart} 
\usepackage{amssymb,amsthm}             
\numberwithin{equation}{section}        
\usepackage{amssymb,amsthm}             

\usepackage[mathscr]{eucal}

\usepackage{url}

\newcommand{\mnote}[1]{}     

\renewcommand{\Re}{\mathbb R}

\newcommand{\genus}{\text{\rm genus}}

\theoremstyle{plain}

\title[On the relation between mathematical and numerical relativity]{On the
  relation between mathematical and numerical relativity}
\author[L. Andersson]{Lars Andersson
} \email{larsa@math.miami.edu} 
\thanks{
Supported in part by the NSF, under 
contract no. DMS 0104402 with the University of Miami.}
\address{Albert Einstein Institute, Am M\"uhlenberg 1, D-14476 Potsdam,
  Germany \and
Department of Mathematics, University of Miami, Coral Gables, FL
33124, USA}

\begin{document}

\begin{abstract}
The large scale binary black hole effort in numerical relativity
has led to an increasing distinction between numerical and mathematical
relativity. This note discusses this situation and gives some examples of
succesful interactions between numerical and mathematical methods is general
relativity.
\end{abstract}


\date{July 2, 2006}

\maketitle


\section{Introduction}
After a lengthy period of fighting various ``monsters'' \cite{Lakatos}, such
as spurious radiation, constraint instabilities, boundary effects, collapse
of the lapse, etc., \mnote{check this list with Luis and Carsten} 
the effort in numerical general relativity directed at modelling mergers of 
binary black
holes is now rapidly entering a phase of ``normal science''. 
Although not all of the monsters have been tamed, 
a number of groups are reporting multiple orbit evolutions and the goal
of providing reliable wave forms in sufficient numbers and of sufficient
accuracy for use in gravitational wave data analysis is in sight. 

The conceptual framework for the numerical work on the binary black hole
(BBH) problem, which arguably has played an absolutely necessary role
as  foundation and stimulus for this work,  has been 
provided by the global picture of spacetimes, 
including singularity theorems, ideas
of cosmic censorship, post-Newtonian and other analytical approximations of
the 2-body problem in general relativity, which 
have been arrived at purely by analytical
and geometric techniques. 
Further, the theoretical analysis
of numerical approximations to solutions of systems of PDE's, the analysis of
the Cauchy problem for the Einstein equations, developments in computer
science concerning parallell processing, all provided essential stepping
stones on the path towards successful BBH simulations. 

Due to the large scale of the effort that goes into the BBH work, the
division of the general relativity community into ``numerical'' and
``theoretical/mathematical'' groups has become pronounced. With this in
mind, it seems natural to ask oneself what grounds there are for future
interactions between these two communities.
On the one hand, one may take the point of view that the ``strong field''
regime of general relativity is going to be the essentially exclusive domain
of numerical general relativity,
the phenomena one is likely to encounter being too complex to be amenable to
mathematical analysis; a consequence of this point of view is the
recommendation to mathematical relativists interested in these aspects of
general relativity
to devote themselves to becoming numerical relativists. 
On the other hand, one may take the point of view that the strong field
regime of general relativity likely contains new phenomena of interest both for
our understanding of the analytical nature of the Einstein equations, as well
as for our understanding of physical reality. 

In the latter point of view, which I am proposing in this note, the relation
between numerical and mathematical general relativity 
\mnote{bring in geometric analysis} 
is similar to that of experimental
mathematics to mathematics, i.e. as a tool for discovering new phenomena,
testing conjectures, and developing a heuristic framework which can be used
in a precise mathematical analysis. In either case, there is a clear need for
an effort to bridge the emerging gap between the two communities. 

\subsection{Numerical experiments and mathematics}
Mathematics has a long history of interaction between computer simulations
and analytical work. Areas where this interaction has been 
prominent are number theory, dynamical systems, and fluid mechanics. The
interaction has provided both the discovery of new phenomena, as well as
proofs of theorems conjectured on the basis of numerical experiments. 

A few examples where 
the interaction between mathematics and computer
simulations has played an important role 
are provided by the accidental discovery in 1963 by Lorentz 
\cite{lorentz} of chaotic
behavior in a system of equations derived from atmospheric
models\footnote{The famous Fermi-Pasta-Ulam experiment of 1955 is perhaps the
non-chaotic counterpart of the Lorentz experiment.}, 
the discovery by Feigenbaum \cite{feigenbaum} of universality in period
doubling bifurcations, the discovery and study of strange attractors in
dynamical systems, and the analysis of fractals including the Mandelbrot set \cite{mandelbrot}.

The proof of the existence of solutions to the Feigenbaum functional
equation was computer based, using rigorous numerical computer techniques \cite{lanford}. 
The existence of strange attractors for the H{\'e}non map \cite{henon} 
was proved by Benedicks
and Carleson \cite{BC}, using analytic techniques. 
The proof was preceeded by a lengthy period of
theoretical work as well as very detailed computer simulations which gave
strong support to the conjectured picture of the attractor and the dynamics
of the H{\'e}non map. The H{\'e}non map was derived as a model for the
Poincar{\'e} map of the Lorentz system. It was recently proved that the
Lorentz system contains a strange attractor \cite{tucker}, 
thus providing a solution to Smale's 14th problem. 
The proof of this fact was again computer based. 

\subsection{Overview of this paper} 
Below, 
in section \ref{sec:success}, 
I shall discuss three examples from general relativity. The first is the
Bianchi IX, or Mixmaster system, an anisotropic homogenous cosmological
model, and in particular modelled by a system of ODE's, see section \ref{sec:mixmaster}. 
The second is the Gowdy $T^2$-symmetric cosmological
model, which is modelled by a 1+1 dimensional system of wave equations, see
section \ref{sec:gowdy}. 
Third, I will discuss critical collapse, see section \ref{sec:critical} 
which was first discovered
during numerical simulations of the collapse of a self-gravitating scalar
field. 

In section \ref{sec:open}, 
I mention some open problems where it seems likely that the interaction of
numerical and analytical techniques will play an important role. In section
\ref{sec:T2}, I discuss general $T^2$ symmetric cosmologies, which provide a
simple model for the full BKL type behavior followed by a few remarks on generic singularities in
section \ref{sec:generic}. 
The next section
\ref{sec:wavemap}, introduces the problem of self-gravitating wave maps and
the $U(1)$ model.  Finally, the stability of the Kerr black hole is
discussed in section \ref{sec:Kerr}. Concluding remarks are given in section
\ref{sec:concluding}. 

\section{Success stories} \label{sec:success}

\subsection{The Mixmaster spatially homogenous cosmology}
\label{sec:mixmaster} 
The Bianchi IX or Mixmaster model is given by restricting the vacuum Einstein
equations to the spatially homogenous case with $S^3$ spatial topology. 
The dynamics of this system was first
discussed in some detail by Misner, see \cite{misner:1993} and references
therein, see also \cite{WE}. 
Misner gave a Hamiltonian analysis which indicated that the system exhibits
``bounces'' interspersed with periods of ``coasting''. The thesis of Chitre
\cite{chitre} 
gave an approximation of the dynamics as a hyperbolic
billiard.
It was
quickly realized that the billiard system is chaotic in a certain sense,
namely it projects to the Gauss map, $x \mapsto \{ \frac{1}{x} \}$, see
\cite{barrow}, which has
been well studied. 

The heuristic picture of the oscillatory, and chaotic, 
asymptotic behavior of the Mixmaster model played a central role in the
proposal of 
Belinski\v{\i}, Khalatnikov, and Lifshitz (BKL) 
concerning the structure of generic singularities for the gravitational
field \cite{bkl70,bkl82}. 
An essential aspect of the BKL proposal is that the dynamics near
typical spatial points is asymptotically ``Mixmaster''-like. In the case of
spacetimes containing stiff matter on the other hand, the asymptotics is
``Kasner''-like, and quiescent. The quiescent behavior also occurs under
certain symmetry conditions, an important example being the Gowdy spacetimes
to be discussed below. Apart from the intrinsic beauty of the Mixmaster
system, the BKL proposal provides one of the main motivations for 
studying the Mixmaster system in detail. 

It should be remarked that the chaotic nature of the Mixmaster dynamics was 
used by
Misner as a basis for the so-called ``chaotic cosmology'' proposal,
 in which
it was argued that  the dynamics of Mixmaster gave a way around the horizon
problem which plauged cosmology during this period (i.e. pre-inflation). 

The full Mixmaster model resisted analysis for a long time, 
in spite of a large
number of papers devoted to this subject. Numerical experiments indicated
that the model has sensitive dependence on initial data, but also revealed
that evolving the system of ODE's describing the Mixmaster dynamics for
sufficiently long times to give useful insights, and with sufficient accuracy
to give reliable results, presented a difficult challenge. It was only with
the work of Berger, Garfinkle and Strasser \cite{BGS}
which made use of symplectic integration techniques and an analytic
approximation that it was possible to
overcome the extremely stiff nature of the system of ODE's for the Mixmaster
model. This numerical work gave strong support for the basic conjectures
concerning the Mixmaster system, and led to a renewed interest within the
mathematical general relativity community in the analysis of the Mixmaster dynamics. The
volumes \cite{HBC} as well as the paper
\cite{rendall:mix} played an important role in spreading the word about
this problem. 

The main conjectures concerning the mixmaster model, including proof of cosmic
censorship in the Bianchi class A models, the oscillatory nature of the
Bianchi IX singularity, as well as the existence of an attractor for the
Bianchi IX system were proved by Ringstr\"om in a series of papers
\cite{ring2000,ring2001}. 
However, in spite of these very important results, many basic and important
questions concerning both the full Mixmaster system, as well as the billiard
approximation, remain open. 

Recent work of Damour, Henneaux, Nicolai and others \cite{DHN} 
have shown that the BKL
conjecture extends in a very interesting way to higher dimensional theories
of gravitation inspired by supergravity theories in D=11 spacetime
dimensions. A formal argument indicates that these models have asymptotic
Mixmaster like behavior, governed by a hyperbolic billiard, determined by the
Weyl chamber of a
certain Kac-Moody Lie algebra. Applying this analysis to 3+1 vacuum gravity
reproduces the Chitre model. These domains occurring in these hyperbolic
billiards are arithmetic, which has interesting consequences for the 
length spectrum of the billiard. 

An important open problem is to understand the relation between the
``Hamiltonian'' approach, developed by Misner-Chitre, and which also is used
in the work of Damour-Henneaux with the scale invariant variables approach
developed by Ellis-Wainwright-Hsu, and which was used in the work of
Ringstrom on Mixmaster. The scale invariant variables formalism has been
generalized to inhomogenous models by Uggla et. al, \cite{uggetal2003}, and
applied to formal and numerical analysis of inhomogenous cosmological models
\cite{AELU,GarfPRL}.

\subsection{The Gowdy $T^2$-symmetric cosmologies} \label{sec:gowdy} 
The cosmological models on $T^3 \times \Re$, with $T^2$ symmetry, and with 
hypersurface
orthogonal Killing fields, the so-called Gowdy model, 
is one of the simplest inhomogenous cosmological models. The Einstein
equations reduce to a system of PDE's on $S^1 \times \Re$, consisting of  a
pair of nonlinear wave equations of wave maps type, and a pair of constraint
equations.  
Eardley, Liang and Sachs \cite{ELS} 
introduced the notion of asymptotically velocity dominated singularities,
to describe the asymptoticall locally Kasner like, non-oscillating behavior
of certain cosmological models. 
We will refer to this
behavior as quiescent. In particular, analysis
showed that one could expect the Gowdy model to exhibit quiescent behavior at
the singularity. 

A programme to study the Gowdy model analytically, with a view towards
proving strong cosmic censorship for this class of models, was initiated by
Moncrief. The methods used included a Hamiltonian analysis, and formal power
series expansions around the singularity. The formal power series expansions
of Grubisic  and Moncrief \cite{GM} supported the idea that a family of Gowdy
spacetimes with ``full degrees of freedom'', i.e. roughly speaking
parametrized by four functions, exhibited  quiescent behavior
at the  singularity.
\mnote{who conjectured this?} 
In the course of this work, an
obstruction to the convergence of the formal power series was
discovered. 
The condition for the consistency of the formal power series
expansions was that the ``asymptotic velocity'' $k$ of the Gowdy spacetime,
satisfies $0 < k < 1$. The term asympotic
velocity has its origin in the fact that 
the evolution of a Gowdy spacetime corresponds to the motion of a
loop in the hyperbolic plane. 
The asymptotic velocity $k(x)$ for $x \in S^1$ 
is defined as the asymptotic hyperbolic velocity of the point with parameter
$x$ on the evolving loop in the hyperbolic plane. \mnote{alternatively as
  part of the energy density} It is a highly nontrivial fact that this
limiting value exists, see \cite{ring:vel}. 

Numerical studies carried out by Berger and Moncrief
\cite{BM93} and later by Berger and Garfinkle \cite{BG98}
gave rise to a good heuristic picture of the asymptotic
dynamics of Gowdy models. In particular, the numerical work showed that Gowdy
spacetimes exhibit sharp features (spikes), which formed and appeared to
persist until the singularity. The spatial scale of the spikes turned out to
be shrinking exponentially fast, and it was therefore impossibly to resolve
these features numerically for more than a limited time. Kichenassamy and 
Rendall \cite{KR98}
showed, using Fuchsian techniques, that the picture developed in
the work on formal expansions could be made rigorous, and in particular that
full parameter families of ``low velocity'' 
Gowdy spacetimes could be constructed  with quiescent singularities. Further,
Rendall and Weaver \cite{RW} used a combination of Fuchsian and solution generating
techniques to construct Gowdy spacetimes containing spikes with arbitrary
prescribed velocity. This allowed one to gain detailed understanding of the
nature of the spikes, and in particular of the nature of the discontinuity 
of the asymptotic velocity at spikes.

These developments gave through the numerical work, a vivid graphical picture
of the dynamics of Gowdy spacetimes, but also established with rigor some of
the fundamental conjectured aspects of Gowdy spacetimes. Based on these
developments, Ringstrom \cite{ring2004,ring2004b,ring2004c} 
was able to analyze the nature of the singularity of
generic Gowdy spacetimes, and in particular give a proof of strong cosmic
censorship for this class of spacetimes. 

\subsection{Critical collapse}
\label{sec:critical} 
Critical behavior in singularity formation was discovered by Matt Choptuik \cite{chop92},
during numerical 
studies of the collapse of self-gravitating scalar fields. He found
that for one-parameter families of initial data, interpolating between data
leading to dispersion and data leading to collapse, data on the borderline
between dispersion and collapse exibited, for a period depending on the
parameter, a discrete self-similar behavior before dispersing or
collapsing. Further, Choptuik found that the rate of divergence from the
self-similar behavior exhibited a ``universal'' behavior, analogous to the
universality discovered by Feigenbaum in connection with period doubling
bifurcations. This work opened up a very rich field of investigation which is
still active. 

The basic principle is now well established through a large
number of numerical experiments and investigations, see the review paper
\cite{gundlach:crit}. A formal analysis
indicates that the ``universal'' behavior mentioned above may be explained in
terms of a linearized analysis around the self-similar critical solution
\cite{evans:coleman,koike:etal}. 
It turns out that the detailed behavior, in particular the rate,
depends on the details of the nonlinearity, or in the case of general
relativity, the matter
model under consideration, but the basic idea of universality within a matter
model, and the above mentioned mechanism for the critical behavior is well
established over a wide range of models. 
Depending on the matter
model, the self-similar behavior may be discrete, continuous, or even in some
cases a mixture of the two types. 
Virtually all numerical 
work on critical behavior has so far been in the
spherically symmetric case. The reason is the extreme demands on numerical
precision presented by the problem. 

By generalizing the notion of critical behavior from general relativity to semilinear wave
equations, Yang-Mills equations, and wave maps equations, Bizon and others
\cite{biz2000}, have  been able in some cases 
to find the explicit form of the first unstable
(self-similar) mode, and thus give an analytic description of the blowup
solutions. They find good agreement with numerical data. 
However, beyond the linearized stablility analysis mentioned above, not many
rigorous results are known for critical behavior for 
the hyperbolic equations mentioned above, including the case of general
relativity.  \mnote{mention Merle} 
This state of affairs should be contrasted with the asymptotic analysis of
singular solutions of semilinear parabolic equations.

In the 2+1 dimensional case, a new phenomenon arises. For wave
maps on (2+1)-dimensional Minkowski space,
with spherical target, there is no self-similar solution. Instead
there is a one-parameter family of static
solutions, and numerical work in the equivariant case shows that this family mediates the
blowup \cite{biz2001}. This is borne out by the proof due to Struwe
\cite{struwe:blow} that in the equivariant case, a
rescaling limit of a blowup solution converges to a harmonic map from $\Re^2$
to $S^2$. Due to this fact, the nature of the critical blowup in 2+1 dimensions
is fundamentally different from the 3+1 dimensional case, and the analysis of
the asympotic rate of concentration of blowup solutions is much more
delicate. Recent work of Rodnianski and Sterbenz \cite{rodster} sheds light
on this question in the general case without symmetries.

\mnote{mention Rodnianski here} \mnote{mention 2+1 self-gravitating
  wave maps -- possible critical behavior in connection with the coupling
  constant} 

The further study of the asymptotic behavior of blowup solutions of
semilinear wave equations as well as the gravitational field, in the
non-spherically symmetric case is one of the important challenges for the
near future. Here it appears likely that a lot of the technology developed during
the course of the BBH work, such as adaptive mesh-refinement, etc. will play
a decisive role. Indeed, the original work by
Choptuik on critical collapse used a version of adaptive mesh refinement in
the spherically symmetric situation. 
 \mnote{mention Choptuik used mesh refinement in his original
  spherically symmetric code} \mnote{Aichelburg-Husa asymptotic mass in
  critical collapse -- what is the situation} 

\section{Open problems} \label{sec:open} 
In this section, I shall briefly indicate some problems which I consider to be
of interest from the point of view of the interaction between numerical and
mathematical work in general relativity and related fields. 
The survey papers
\cite{And:sur,Rend:sur} provide general
references for many of the problems mentioned below.

The problems I will mention are not exactly coincident with the ``forefront''
of numerical relativity, and may by some workers in that field be considered
as simple problems, not worthy of their attention. 
There are several reasons for this. One is that the
development of the mathematical theory of relativity is to a large extent
lagging behing the exploratory and goal oriented work being performed
within numerical relativity. Further, in order to provide reliable 
insights into the nonlinear problems under consideration, the numerical
experiments must necessarily be carried out to a high degree of
accuracy. This level of precision is so far not available in general in 
the 3+1 or even
2+1 dimensional numerical evolutions, in particular not in the strong field
regime where many of the phenomena of interest take place. In particular, a
serious numerical study of the asymptotic behavior at cosmological and other
types of singularities, with a view to better understanding the BKL proposal
in general relativity, as well as the asymptotic behavior of blowup solutions
of geometric wave equations, without symmetry assumptions, is likely to be at
least as challenging as the BBH problem.

\subsection{General $T^2$-symmetric comologies} \label{sec:T2} 
The full $T^2$ symmetric model on $T^3 \times \Re$, 
without the condition the the Killing fields
be surface orthogonal exhibits oscillatory behavior at the singularity. While
this has not been rigorously established, this is indicated by formal and
numerical studies. The formal work includes the analysis of the silent
boundary due to Uggla et al \cite{uggetal2003}.
There are formal and 
numerical studies using both the metric formulation by Berger et al.
\cite{ber2001} and the scale invariant formulation by Andersson et al.
\cite{AELU}. The last mentioned work gives numerical support to the silent
boundary picture for the case of $T^2$-symmetric cosmologies. 
The review by Berger \cite{ber2002} provides a 
general reference on the numerical investigation of
spacetime singularities. 
The numerical studies lend support to the BKL proposal on
the nature of generic cosmological singularities, and also indicate some
new dynamical features of the $T^2$ singularity. These new features can be
interpreted as spikes. However in contrast to the Gowdy case, where the
asymptotic velocity at the spike is a constant, 
the spikes in $T^2$ are dynamical features, which according to the numerical
experiments \cite{AELU} 
exhibit a simple dynamics, closely related to the asymptotic
billiard for the silent boundary system for $T^2$.

\subsection{Singularities in generic cosmologies} \label{sec:generic} 
The $U(1)$ model has been studied in the spatially compact case by
Choquet-Bruhat and Moncrief \cite{ChBM,ChB}. They proved global
existence in the expanding direction for small data on spacetimes with
topology $\Sigma \times \Re$ with $\genus(\Sigma) > 1$. For the polarized
case, one has a self-gravitating scalar field. In the polarized case, one
expects to have quiescent behavior at the singularity, a full parameter
family of such solutions was constructed \mnote{in the half-polarized case}
by Choquet-Bruhat, Isenberg and Moncrief 
\cite{IM,ChBIM}. For the full $U(1)$ model, one expects an
oscillatory singularity, as in the full $T^2$ case. Numerical and analytical 
work of Berger
and Moncrief \cite{BM2000,BM2000b} give support to this picture. 
For this model, as for
the $T^2$ model, the BKL proposal, and in particular, the silent boundary
proposal of Uggla et al \cite{uggetal2003} provides a heuristic picture of
the dynamical behavior that one expects to see. 

In fact, the scale invariant
variables introduced by Uggla et al provides, with
for example CMC time gauge, a well posed elliptic-hyperbolic system, which
can be used to model the dynamical behavior at the singularity. Some
preliminary numerical experiments using a CMC code have been carried out.
Even for the polarized case in the 2+1 dimensions
the need for adaptive codes is apparent. The
oscillatory nature of the singularity means 
that spatial structure is created at small scales. This will make it
impossible to produce even somewhat realistic evolutions of the full $U(1)$
model without using an adaptive code. See however the recent work by
Garfinkle \cite{GarfPRL} for some
numerical experiments in the 3+1 case. For these, even though they reproduce
the heuristic picture derived from the silent boundary conjecture, 
the accuracy is too low to
provide reliable information. Earlier work by Garfinkle \cite{Gar2002} 
on cosmological
singularities in 
self-gravitating scalar field model in 3+1 dimensions made use of spacetime
harmonic (or wave) coordinates. The fact that this experiment, which for
essentially the first time made use of spacetime harmonic coordinates for a
numerical relativity code was successful,
has later had a significant influence on current work on the BBH problem. 
The self-gravitating scalar field model is known to have large families of
data which give rise to quiescent singularities \cite{AR2001}. 
 
\subsection{Self-gravitating wave maps and $U(1)$} \label{sec:wavemap} 
\mnote{mention Aichelburg, Bizon et al on the 3+1 dimensional case}  
Vacuum 3+1 dimensional gravity with a spatial $U(1)$ action gives, after a
Kaluza-Klein reduction, a self-gravitating wave-maps model in 2+1
dimensions, with hyperbolic target space. Further imposing on this model a
rotational symmetry, i.e. another spatial $U(1)$ action, which acts
equivariantly, results in an equivariant self-gravitating 2+1 dimensional
wave map with hyperbolic target. The equivariant $U(1)$ action does not
correspond to a Killing field in the 3+1 dimensional picture. In this case it
is natural to impose asymptotic flatness for the 2+1 dimensional spacetime
\cite{ashvar}. If this is done, turning off the gravitational
interaction gives a flat space wave maps model with hyperbolic target. 

According to standard conjectures, the flat
space 2+1 dimensional wave map with hyperbolic target is expected to be
well posed in energy norm, see \cite{tao2004}, 
and it is therefore reasonable to expect that also the
self-gravitating version of this model is globally well-posed. 
For the 2+1 dimensional wave maps model with spherical target, on the other
hand, numerical experiments indicate that one has blowups for large data, see
section \ref{sec:critical} above. 


Based on the idea that the blowup in the wave maps model with spherical
target is mediated by static solutions, one may argue that in the
self-gravitating case, 
one cannot have blowups
for sufficiently large values of the coupling constant. The reason for this
is that the energy balance between the gravatational field and the wave maps
field does not leave enough energy for the wave maps field to produce the
static solutions which mediate blowup. \mnote{Bogomolny bounds} 
Some numerical experiments have been
carried out \mnote{add cite} which support this picture. It would be of
great interest to have detailed numerical experiments in this situation. The
full 2+1 dimensional version of the self-gravitating wave maps problem is 
very challenging both numerically and
theoretically. 

\subsection{Stability of Kerr} \label{sec:Kerr} 
As mentioned above, the understanding of the structure of full 3+1
dimensional cosmological singularities and the strong cosmic censorship
represents a major challenge to the numerical and mathematical relativity
community. However, the stability of the Kerr black hole is perhaps closer to
the type of problems which occupy most of the attention of current work in
numerical relativity. 
%
As is well known, according to the cosmic
censorship picture, the end state of the evolution of an asymptotically flat
data set is a single Kerr black hole.
A proof
of the nonlinear stability of Kerr would provide an important step
towards a proof of this far-reaching conjecture. 

The nonlinear stability of Minkowski space was proved
by Christodoulou and Klainerman, see \cite{ChK}, see also \cite{friedrich:complete} 
for an earlier partial result, using a conformally regular form of the Einstein
equations. 
For quasilinear wave equations which satisfy the so-called null condition of
Christodoulou \cite{Chrnull}, global existence for sufficiently small data is known to
hold in dimension $n+1$, $n \geq 3$. 
It is a very important fact that the 
Einstein equations do not satisfy the classical null condition
\cite{ChBnull}.

The
proof of Christodoulou and Klainerman relied upon detailed and rather delicate 
estimates of higher order Bel-Robinson energies, using a combination of
techniques. The geometry of certain null foliations was
studied, exploiting the transport equations for geometric data along null
rays. Further, a variant of the vector fields method of Klainerman
\cite{Kl:weight} was used. 

The vector fields method was developed to prove decay estimates for solutions
of wave equations, and requires at least approximate symmetries of the
background solution. The method has been used by Klainerman and Rodnianski in
a micro-local setting in order to prove well-posedness for the Einstein
equations with rough data. 
%
%
The method of Christodoulou and Klainerman has later, in a series of papers
by Nicolo and Klainerman \cite{KN:book,KN2003} 
been shown to yield the correct peeling behavior at
null infinity, which is expected from the Penrose picture. 

Recently, a substantially simpler proof of the nonlinear stability of
Minkowski space was given by Lindblad and Rodnianski \cite{LR}. Their proof
relies upon the so-called weak null condition. 

Neither of the above mentioned techniques generalize easily to the case of a
non-flat backgrund solution. One serious problem is that the light cones
in a black hole spacetime differ by a logarithmic term 
from those in Minkowski space. Further,
due to the presence of the horizon, and in particular the ergo region, one
has different types of decay behavior in the region close to the
black hole and in the asymptotic region. 

Several natural problems arise in this context. The decay of scalar field on
Schwarzschild and Kerr backgrounds, in particular the behavior at the horizon
(Price law) is a natural starting point. Several recent papers have studied
this problem \cite{DR2005,FKSY} and for the case of a Schwarzschild
background the estimates agree with the conjectured Price law behavior. 

While some mathematical results on for example the decay of scalar fields on
Schwarzschild and Kerr backgrounds are available, the techniques used to
prove these are intimately tied to the symmetries of the background, and make
heavy use of spherical harmonics expansions. Therefore these proofs do not
directly generalize to spacetimes which are close in a suitable sense to
Kerr. Further, one could even say that we don't have a good notion of what
``close to Kerr'' actually means.

Thus, the problem of stability of Kerr opens up a natural arena for the
interaction of numerical and mathematical relativity. The aspects of
this problem where numerical experiments may be able to provide crucial
insights include the asymptotic decay behavior of the gravitational field and
matter fields near the horizon. A question closely related to this, and of
direct relevance for numerical work, is the asymptotic behavior of dynamical
horizons \cite{AK,AG,AMS,SKB}. 
It is not unlikely that a good understanding of the asympotic
geometry of dynamical horizons near timelike infinity will play a crucial
role in the global analysis of black hole spacetimes. 

In the far region and intermediate region, one expects linear effects to
dominate and here there is a lot of information available from systematic
post-Newtonian calculations. It is of interest to compare this to the results
of numerical simulations, and a great deal of work in this direction is
already being carried out in the context of the BBH programme.


\section{Concluding remark} \label{sec:concluding} 
This note represents a personal view and the
rather incomplete discussion here leaves out very large areas of numerical relativity and
numerical geometric analysis, including Ricci flow, heat flow, higher
dimensional general relativity models, including black strings 
and other areas which are being worked on intensely. Further, the asymptotic
behavior of cosmologies in the expanding
direction, which has not been discussed here,  
provides interesting open questions, 
which can be fruitfully studied using
numerical techniques.  

\subsection*{Acknowledgements} 
This note is loosely based on a talk given during the Newton Institute
programme on Global Problems in Mathematical Relativity, fall of 2005. 
%
I am grateful to the organizers of the GMR programme 
for inviting me to the Newton Institute and
to the Newton institute for the excellent working environment. I would also
like to thank the editors 
of the special issue on numerical general
relativity, for the invitation to contribute an article.

\def\cprime{$'$}
\providecommand{\bysame}{\leavevmode\hbox to3em{\hrulefill}\thinspace}
\providecommand{\MR}{\relax\ifhmode\unskip\space\fi MR }
\providecommand{\MRhref}[2]{%
  \href{http://www.ams.org/mathscinet-getitem?mr=#1}{#2}
}
\providecommand{\href}[2]{#2}

\end{document}